\documentstyle[12pt,epsfig]{article}
\begin{document}

\begin{titlepage}
\rightline{\large Nov 2011}
\vskip 2cm
\centerline{\Large \bf
Mirror dark matter cosmology - }
\vskip 0.5cm
\centerline{\Large \bf
predictions for $N_{eff} [CMB]$ and $N_{eff} [BBN]$}

\vskip 2.2cm
\centerline{\large R. Foot\footnote{
E-mail address: rfoot@unimelb.edu.au}}

\vskip 0.7cm
\centerline{\it ARC Centre of Excellence for Particle Physics at the Terascale,}
\centerline{\it School of Physics, University of Melbourne,}
\centerline{\it Victoria 3010 Australia}
\vskip 2cm
\noindent
Mirror dark matter interacting with ordinary matter via photon-mirror photon kinetic mixing 
can explain the DAMA, CoGeNT and CRESST-II direct detection experiments. This explanation 
requires kinetic mixing of strength $\epsilon \sim 10^{-9}$. Such kinetic mixing will have 
important implications for early Universe cosmology.  We calculate the additional 
relativistic energy density at recombination,  $\delta N_{eff} [CMB]$.
We also calculate the effects for big bang nucleosynthesis, $\delta N_{eff} [BBN]$.
Current hints that both
$\delta N_{eff} [CMB]$ and $\delta N_{eff} [BBN]$ are non-zero and positive
can be accommodated within this framework if $\epsilon \approx {\rm few} \times 10^{-9}$.
In the near future, measurements from the Planck mission will either
confirm these hints or constrain $\epsilon \stackrel{<}{\sim} 10^{-9}$.

 \end{titlepage}

The long standing DAMA annual modulation signal\cite{dama}, along with more recent data from
the CoGeNT\cite{cogent} and CRESST-II\cite{cresst} experiments can be explained within a simple hidden sector 
framework\cite{foot11,talk} (see also ref.\cite{foot69,footold} for earlier studies).  
The requirement is for a hidden sector with an unbroken $U(1)'$ 
interaction kinetically mixed with the standard $U(1)_Y$ gauge field.
Such a framework can explain all of the direct detection results, both positive and negative, 
provided that the hidden sector contains two or more stable particles.
Mirror dark matter\cite{flv} appears to be the most well motivated and also 
constrained example of such a theory (see ref.\cite{review}
for reviews and more comprehensive references).
In the mirror dark matter theory, the hidden sector is exactly isomorphic to the 
ordinary sector except the chiralities are interchanged. In this case an exact unbroken mirror
symmetry can be defined which interchanges each ordinary particle with a mirror partner,
as well as mapping $x, t \to -x, t$\cite{flv}. The latter property allows mirror symmetry
to be identified with space-time parity which provides an interesting theoretical motivation for
such a theory\footnote{There are also interesting models with mirror symmetry spontaneously 
broken, see e.g.\cite{spon,spon2}.}.

In this theory dark matter consists of a spectrum of stable mirror particles 
of known masses (e$'$, H$'$, He$'$, O$'$, ...) which interact with the
ordinary particles via gravity and renormalizable photon-mirror photon kinetic mixing:\cite{he}
\footnote{Other interactions are possible including a Higgs-mirror Higgs 
coupling $\lambda \phi^\dagger \phi \phi'^\dagger \phi'$, 
which can lead to novel effects for Higgs physics that can be probed
at the LHC\cite{mirrorhiggs}.}
\begin{eqnarray}
{\cal L}_{mix} = \frac{\epsilon}{2} F^{\mu \nu} F'_{\mu \nu}
\label{kine}
\end{eqnarray}
where $F_{\mu \nu}$ is the ordinary photon
field strength tensor, and $F'_{\mu \nu}$ is the field strength tensor for the mirror photon.
This interaction enables mirror charged particles, such as e$'$, p$'$, to couple to
ordinary photons with electric charge $\epsilon e$ \cite{holdom}.

Studies have shown\cite{lssstudies} that mirror dark matter can provide a successful account of 
the Large Scale Structure (LSS) of the Universe which can mimic cold dark matter
or warm dark matter, depending on initial conditions.  Mirror dark matter, though, has quite
distinctive properties on smaller scales. Observations suggest that 
a) In galaxy clusters or at least in some of them, mirror dark matter should be predominately
confined within galactic halos to account for the observations of the Bullet cluster\cite{bullet}
(c.f. \cite{zurab}). 
b) Within galactic halos, limits on the  MACHO fraction\cite{macholimit} 
suggest that mirror dark matter is mainly (at least $\sim 50\%$)
in gaseous form - i.e. an ionized plasma comprising e$'$, H$'$, He$'$, O$'$, .... 
A substantial MACHO fraction is also possible\cite{macho}.
Mirror dark matter is dissipative, which indicates that a substantial
heat source should exist to replace the energy lost due to radiative cooling. 
It turns out that ordinary supernova can supply the required
energy to stabilize the mirror dark matter halo provided that $\epsilon \sim 10^{-9}$\cite{fv}.
This heat source is anisotropic in spiral galaxies (given that the 
supernova originate predominately from the disk) which might
explain the deviations from perfect spherical halos which are necessary to agree
with observations.
c) The self-interactions 
of mirror dark matter within galactic halos might help explain the 
inferred cored dark matter distribution - a long standing puzzle of the standard
cold dark matter paradigm.

The gaseous component of the halo is presumed to be composed of an ionized 
plasma comprising a spectrum of components e$'$, H$'$, He$'$, O$'$,..... These components
can potentially be detected in direct detection experiments via the kinetic
mixing interaction, Eq.(\ref{kine}), which induces spin-independent elastic Rutherford
scattering.
The H$'$, He$'$ components, which are naively expected to be dominant, are however very 
light.  In fact these components are so light that they are unable to explain 
the DAMA, CoGeNT or CRESST-II signals.
The experiment most sensitive to the He$'$ component turns out to be the earlier
CRESST/Sapphire experiment\cite{cresst1}.
That experiment can be used to infer the bound:\cite{foot69}
\begin{eqnarray}
\epsilon \sqrt{\xi_{He'}} \stackrel{<}{\sim} 3\times 10^{-9}.
\label{bla33}
\end{eqnarray}
The on-going Texono experiment\cite{texono} at Jin-Ping underground laboratory 
is expected to provide a much more sensitive probe of the He$'$
component in the near future.
The remaining heavier components, are for simplicity assumed to be composed of a `dominant' metal 
component, $A'$.
The direct detection experiments, DAMA, CoGeNT and CRESST-II can be explained by
the $A'$ component if\cite{foot11,talk}:
\begin{eqnarray}
\epsilon \sqrt{\xi_{A'}} &\sim & {\rm few} \times 10^{-10}, \nonumber \\
\frac{m_{A'}}{m_p} &\approx & 16 - 56
\label{bla}
\end{eqnarray}
where $\xi_{A'} \equiv n_{A'}m_{A'}/(0.3 \ {\rm GeV/cm}^3)$ is the halo mass fraction of the species
$A'$ and $m_p$ is the proton mass. 
Combining Eq.(\ref{bla33}) and Eq.(\ref{bla}), we obtain a lower limit on
the ratio $\xi_{A'}/\xi_{He'} \stackrel{>}{\sim} 10^{-2}$ at the Earth's location in the halo.
This appears to be a relatively high metal component, and might be due to rapid
mirror star formation and evolution at an early epoch\cite{mirrorstarstudy}. It might also be possible
to produce the metal component in the early Universe, although this is disfavoured in
the simplest scenarios with high reheating temperature\cite{paolo2}.
Allowing for possible uncertainties in $\xi_{A'}$, Eq.(\ref{bla}) 
suggests that $\epsilon$ should be in the range:
\begin{eqnarray}
10^{-10} \stackrel{<}{\sim} \epsilon \stackrel{<}{\sim} 10^{-8}
\ .
\label{stupidh}
\end{eqnarray}
This range of $\epsilon$ is consistent with the
direct laboratory limits, the most stringent of which arises 
from invisible decays of orthopositronium\cite{gla,ortho}, $\epsilon \stackrel{<}{\sim} 1.5 \times 10^{-7}$.
An important proposal exists\cite{ortho2} for a more sensitive orthopositronium experiment which
can cover much of the $\epsilon$ range of interest, Eq.(\ref{stupidh}).
\footnote{There are a number of other possible implications of kinetic mixing, for example 
if the solar system contains mirror matter space-bodies\cite{mitra}, see 
also the review\cite{review3}.}

In the context of early Universe cosmology,
the kinetic mixing can induce processes which transfer energy into the mirror sector.
This will lead to various
modifications to big bang nucleosynthesis (BBN), Cosmic Microwave Background anisotropies (CMB) 
and also Large Scale Structure (LSS). 
This allows values
of $\epsilon \stackrel{>}{\sim} 10^{-9}$ to be probed in early Universe
cosmology.   

Recall that the relativistic energy component at recombination can be parameterized in terms of the
effective number of neutrino species, $N_{eff} [CMB]$, by:
\begin{eqnarray}
\rho_{rad} = \left( 1 + {7 \over 8} \left[ {4 \over 11}\right]^{4/3} N_{eff} [CMB]\right) \rho_\gamma
\label{5yeah}
\end{eqnarray} 
where $\rho_\gamma$ is the energy density of the CMB photons. In the standard electroweak
theory, $N_{eff} \simeq 3.046$ (the difference from 3 being due to slight heating of the
neutrinos from $\bar e e$ annihilation)\cite{mango}. 
Currently there are some interesting hints that $\delta N_{eff} [CMB] \equiv N_{eff} [CMB] - 3.046$
is in fact non-zero and positive within the usual $\Lambda$CDM framework. For example,
the Wilkinson Microwave Anisotropy Probe (WMAP) obtained $N_{eff} [CMB] = 4.34 \pm 0.87$ (1 $\sigma$ C.L.)
\cite{wmap}.
Observations with the Atacama Cosmology Telescope suggest $N_{eff} [CMB] = 4.56 \pm 0.75$ (1 $\sigma$ C.L)\cite{ata}.
Also, results of the South Pole Telescope combined with WMAP7 suggest $N_{eff} [CMB] = 3.86 \pm 0.42$
(1 $\sigma$ C.L.)\cite{stp}.
A recent analysis\cite{comb} combining the data finds that:
\begin{eqnarray}
N_{eff} [CMB] = 4.08^{+0.71}_{-0.68}\ \ {\rm at}\ \ 95\% \ C.L.
\label{455}
\end{eqnarray}
An effective neutrino number can also be defined for BBN, 
and it is worth noting that two recent 
measurements\cite{bbn1,bbn2} of $Y_p$ suggest $N_{eff} [BBN] > 3$.
These results have motivated a number of proposals for new physics, including models with
sterile neutrinos\cite{models}. 
Importantly, high precision results from the Planck mission should be able to 
confirm a non-zero $\delta N_{eff} [CMB]$ to within a precision of around 0.2\cite{planckexp} in
the near future. In the meantime, it is timely to work out the predictions for $N_{eff} [CMB]$ and $N_{eff} [BBN]$
in the mirror dark matter model.

In the $\epsilon \to 0$ limit, the ordinary and mirror particles decouple from each other, and 
in general they may have different temperatures, $T, \ T'$.
It will be useful to define the temperatures: $T_\gamma$ [$T'_\gamma$] for the temperature of the ordinary 
[mirror] photons and $T_\nu$ [$T'_\nu$] is the temperature of the ordinary [mirror] neutrinos.
In this study, we assume the initial conditions $T'_\gamma , T'_{\nu}  \ll T_\gamma = T_\nu$ 
due to some physics at early times. 
Asymmetric reheating within inflationary scenarios is one possibility\cite{asym}. 
If we assume that only the ordinary matter
is reheated after inflation, then we have the initial condition 
$T'_\gamma , T'_\nu \simeq 0$ and $T_\gamma = T_\nu = T_{RH}$.
With these initial conditions it is always safe to ignore $T'_\nu$ since $T'_\nu \ll T'_\gamma$ in the period of interest. 
This is because
$T'_{\gamma}/T_\gamma$ generation occurs mainly in the low temperature region, $T \stackrel{<}{\sim}$ few MeV,
where the mirror weak interaction rate is always much less than the expansion rate: $G_F^2 T'^5 \ll T^2/m_{pl}$.

For $T_\gamma < 100$ MeV, $\bar e e \to \bar e' e'$ is the dominant process which transfers energy into the 
mirror sector.
This process will not only generate
$T'_\gamma$ but will also slightly reduce $T_\gamma$. This energy transfer happens predominately after
the kinetic decoupling of the neutrinos (i.e. for $T_\gamma \stackrel{<}{\sim} 3 $ MeV), 
so the effect is to slightly decrease $T_\gamma/T_\nu$.
Therefore, at recombination the energy density of neutrinos will be larger than it would 
otherwise have 
been if there were no kinetic mixing. This effect will contribute 
to $\delta N_{eff} [CMB]$ along 
with the more obvious contribution to $\delta N_{eff} [CMB]$ from the increase in energy density 
due to the production of the mirror photons.
We will see that these two contributions are numerically similar in magnitude.

%

The cross-section for the process $\bar e e \to \bar e' e'$ is:
\begin{eqnarray}
\sigma = {4\pi \over 3} \alpha^2 \epsilon^2 {1 \over s^3} (s + 2m_e^2)^2\ 
\end{eqnarray}
where $s$ is the usual Mandelstam variable and $m_e$ is the electron mass.
The energy transfer to the mirror sector within a co-moving volume, $R^3$, per unit time is:
\begin{eqnarray}
{dQ \over  dt} =
R^3 n_{e^+} n_{e^-} \langle \sigma v_{M\o l} {\cal E} \rangle
\end{eqnarray}
where ${\cal E}$ is the energy transferred in the process $\bar e e \to \bar e' e'$,
$v_{m\o l}$ is the M\o ller velocity (see e.g. ref.\cite{gondolo}),
and $n_{e^-} \simeq n_{e^+}$ is the number density of electrons:
\begin{eqnarray}
n_{e^-} = {1 \over \pi^2} \int^{\infty}_{m_e} { \sqrt{E^2 - m_e^2} E
\over  1 + exp(E/T_\gamma) }  \ dE \ .
\label{bla76}
\end{eqnarray}
The quantity $\langle \sigma v_{M\o l} {\cal E} \rangle$ is given by\cite{paolo1} (see also ref.\cite{gondolo}):
\begin{eqnarray}
\langle \sigma v_{M\o l} {\cal E} \rangle =
{\omega \over 8m^4_e T_{\gamma}^2 [K_2 (m_e/T_\gamma)]^2} \int_{4m_e^2}^{\infty} ds \sigma (s -
4m_e^2)
\sqrt{s} \int_{\sqrt{s}}^{\infty} dE_+ e^{-E_+/T_\gamma} E_+ \sqrt{{E_+^2 \over s}
- 1} 
\nonumber
\\
\
\label{bla25}
\end{eqnarray}
where $K_2$ is the modified Bessel function of order 2. The quantity $\omega \approx 0.8$
takes into account various approximations in deriving Eq.(\ref{bla25}) such as 
replacing the exact Fermi-Dirac distribution
of $\bar e, e$ with the simpler Maxwellian one\cite{paolo1}.

Energy is transferred from the ordinary particles to the mirror particles. 
The second law of thermodynamics can be used to work out the change in
entropy of the ordinary particles:
\begin{eqnarray}
dS = {-dQ \over T_\gamma} \ .
\end{eqnarray}
To a very good approximation, we can neglect the contribution of the neutrinos
to $dS$ since energy is mostly transferred to the mirror sector after neutrino kinetic decoupling 
(i.e. for $T_\gamma \stackrel{<}{\sim} 3$ MeV).
Thus, $R \propto 1/T_\nu$, and the above equation has the form:
\begin{eqnarray}
{d \over dt} \left[ { (\rho_\gamma + p_\gamma + \rho_e + p_e)R^3 \over T_\gamma}\right] 
= - {n_{e^-} n_{e^+} \langle \sigma v_{mol} {\cal E} \rangle R^3 \over T_\gamma}
\label{1yay}
\end{eqnarray}
where
\begin{eqnarray}
\rho_\gamma &=& {\pi^2 \over 15} T_{\gamma}^4 \nonumber \\
p_{\gamma} &=&  {\rho_\gamma \over 3} \nonumber \\
\rho_e &=& {2T_\gamma^4 \over \pi^2} \int^{\infty}_{x} {(u^2 - x^2)^{1/2} u^2 
\over 1 + e^u} \ du \nonumber \\
p_e &=& {2T_\gamma^4 \over 3\pi^2} \int^{\infty}_{x} {(u^2 - x^2)^{3/2} 
\over 1 + e^u} \ du 
\end{eqnarray}
and $x = m_e/T_\gamma$.
The entropy gained by the mirror particles is:
\begin{eqnarray}
dS = {dQ \over T'_\gamma} \ .
\end{eqnarray}
This equation implies that
\begin{eqnarray}
{d \over dt} \left[ {(\rho'_\gamma + p'_\gamma + \rho'_e + p'_e)R^3 \over T'_\gamma}\right] 
=  {n_{e^-} n_{e^+} \langle \sigma v_{mol} {\cal E} \rangle R^3 
\over T'_\gamma}
\label{2yay}
\end{eqnarray}
where
\begin{eqnarray}
\rho'_\gamma &=& {\pi^2 \over 15} {T_{\gamma}'}^4 \nonumber \\
p'_{\gamma} &=& {{\rho'_\gamma} \over 3} \nonumber \\
{\rho_e'} &=& {2{T_{\gamma}'}^4 \over \pi^2} \int^{\infty}_{x'} {(u^2 - {x'}^2)^{1/2} u^2 
\over 1 + e^u} \ du \nonumber \\
p_e' &=& {2{T_{\gamma}'}^4 \over 3\pi^2} \int^{\infty}_{x'} {(u^2 - {x'}^2)^{3/2} 
\over 1 + e^u} \ du 
\end{eqnarray}
and $x' = m_e/T'_\gamma$.
The Friedmann equation reads:
\begin{eqnarray}
\left({\stackrel{.}{R} \over R}\right)^2 &=& {8\pi \over 3m_{pl}^2}\left[ \rho_\gamma + \rho_e + \rho_\nu + 
\rho'_\gamma + \rho'_e \right] \ 
\label{3yay}
\end{eqnarray}
where $m_{pl} \simeq 1.22 \times 10^{22}$ MeV is the Planck mass.
The energy density of the neutrinos is given by the standard value, $\rho_\nu = {7\pi^2 \over 40} T_\nu^4$.
Eq.(\ref{1yay},\ref{2yay}) and Eq.(\ref{3yay}) form a closed system which can be solved to 
give the evolution
of $T_\gamma$, $T_\nu$ and $T'_\gamma$.  By way of example, the result for $\epsilon =  10^{-9}$
is shown in figure 1.

\centerline{\epsfig{file=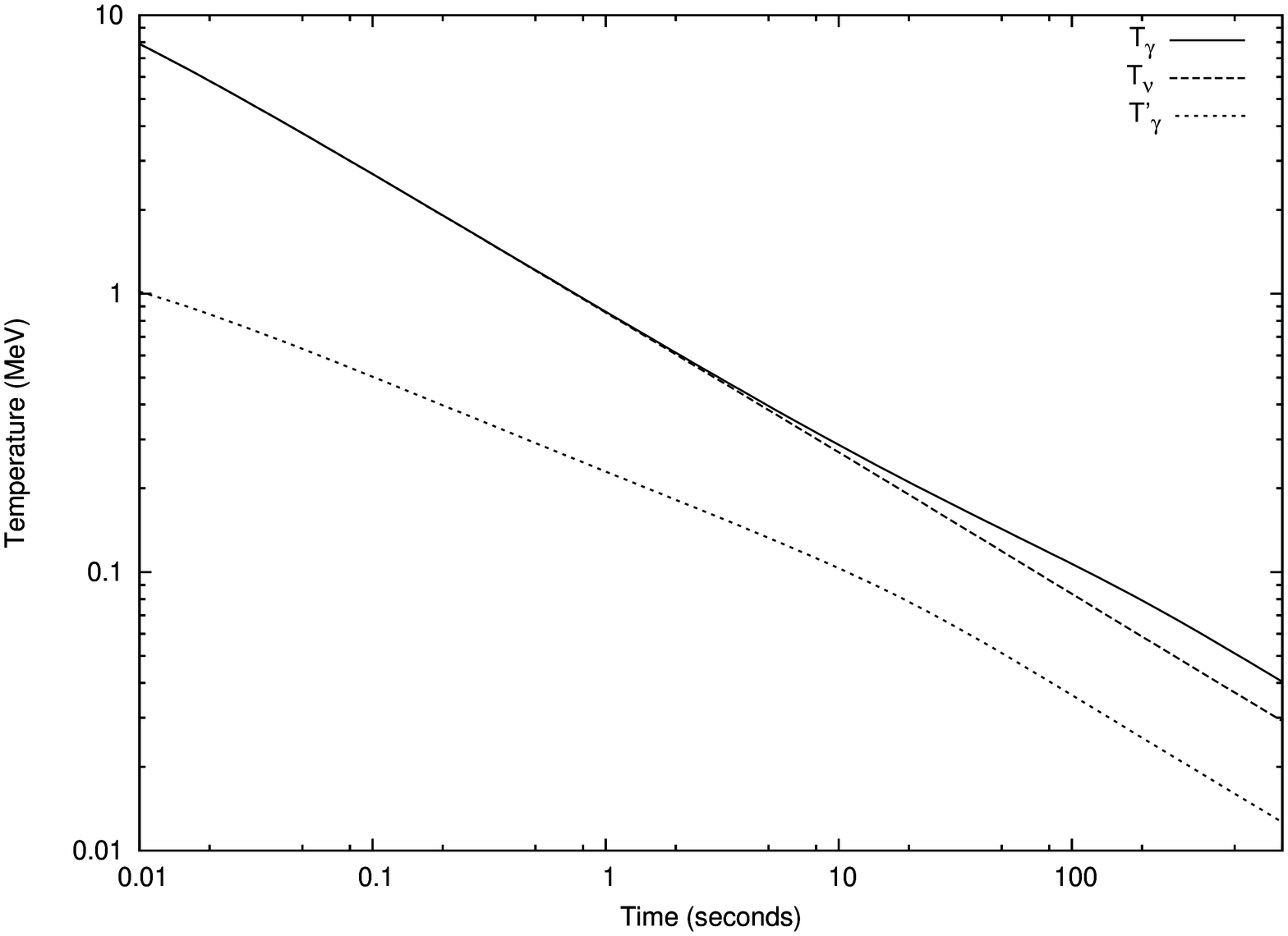,angle=270,width=13.0cm}}
\vskip 0.3cm
\noindent
{\small
Figure 1: Evolution of $T_\gamma$ (solid line), $T_\nu$ (dashed line) and $T'_\gamma$ (dotted line) for
$\epsilon = 10^{-9}$.
}

\vskip 0.8cm
\noindent
We find numerically that $T'_\gamma/T_\gamma$ evolves to a constant which
satisfies:
\begin{eqnarray}
{T'_\gamma \over T_\gamma} \simeq 0.31\left( {\epsilon \over 10^{-9}}\right)^{1/2}
\label{constant}
\end{eqnarray}
which is consistent with the results of the earlier study\cite{paolo1}.

The energy density of the relativistic components 
at recombination can be expressed 
in terms of an effective neutrino number, Eq.(\ref{5yeah}).
The change in this effective neutrino number due to the kinetic mixing effect, 
$\delta N_{eff} [CMB]$, can be computed as follows:
\begin{eqnarray}
\delta N_{eff} [CMB] = 3 \left( \left[ {T_\nu (\epsilon) \over T_\nu (\epsilon = 0)}\right]^4  \ - \ 1\right) \
+ \  {8 \over 7}\left( {T'_\gamma (\epsilon) \over T_\nu (\epsilon =0)}\right)^4  
\end{eqnarray}
where the temperatures are evaluated at the time when photon decoupling occurs, 
i.e. when $T_\gamma = T_{dec} = 0.26$ eV.
The first term on the right-hand side, $\delta N_{eff}^{a}$, is the contribution due to the photon cooling, which
increases the relative energy density of the neutrinos.
The second term, $\delta N_{eff}^{b}$, is the contribution to the energy density from the mirror photons.
In figure 2 we plot $\delta N_{eff} [CMB]$ versus $\epsilon$, showing also the 
separate contributions $\delta N_{eff}^{a}$ and $\delta N_{eff}^{b}$. As remarked earlier [see 
the discussion around Eq.(\ref{455})] there are current hints that
$\delta N_{eff} [CMB]$ is non-zero and positive.  Indeed, the result
of the combined analysis\cite{comb},
Eq.(\ref{455}), provides an interesting hint that $\epsilon \approx few  \times 10^{-9}$.

\vskip 0.5cm
\centerline{\epsfig{file=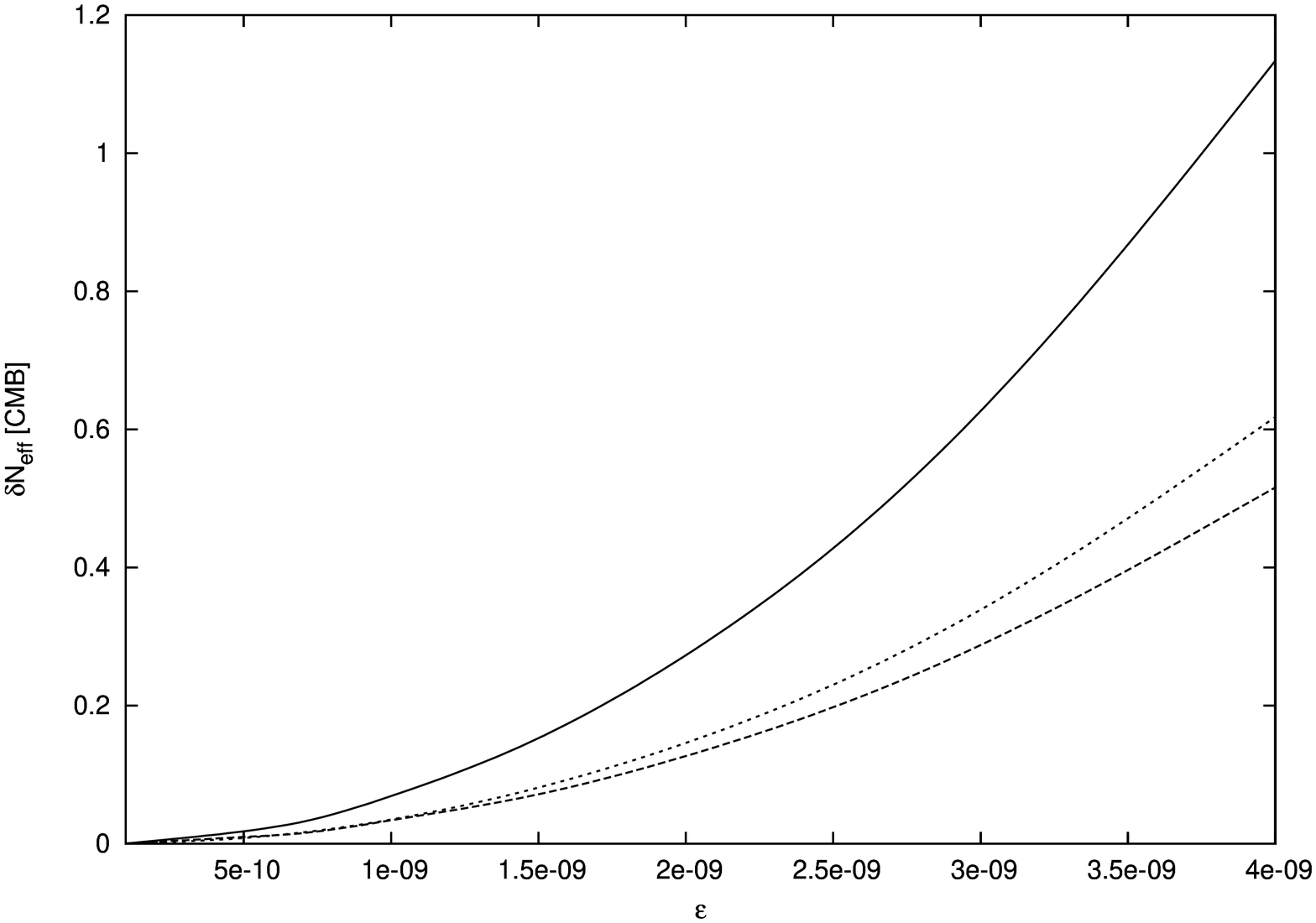,angle=270,width=13.5cm}}
\vskip 0.3cm
\noindent
{\small
Figure 2: $\delta N_{eff} [CMB]$ versus $\epsilon$ (solid line). 
The separate contributions, discussed in the text, $\delta N_{eff}^{a}$ (dashed line)
and $\delta N_{eff}^{b}$ (dotted line) are also shown.
}

\vskip 0.8cm
\noindent

Previous studies\cite{lssstudies} based on successful LSS have suggested that $T'_\gamma/T_\gamma$
should obey an upper limit of around $0.3$.  As pointed out previously\cite{paolo1}, Eq.(\ref{constant})
then suggests an upper limit on
$\epsilon$ of around $\epsilon \stackrel{<}{\sim} 10^{-9}$.
However due to the complexity of modeling the effects of mirror dark matter for LSS
there is a significant uncertainty (maybe even a factor of two or so) in the limit 
on $T'_\gamma/T_\gamma$ from the existing studies based on LSS considerations,
and thus a more conservative upper limit on $\epsilon$ is likely to be around $\epsilon \stackrel{<}{\sim} 4
\times 10^{-9}$.
Obviously if the Planck mission confirms the current hints that $\delta N_{eff} [CMB] > 0$,
then this would warrant a careful re-analysis of the LSS implications of the mirror dark matter
theory.
 
The transfer of energy into the mirror sector also affects BBN as first pointed out in ref.\cite{cg}.
The effect on the primordial Helium abundance can easily be obtained
by computing $Y_p$ in the usual way, but with the $T_\gamma, T_\nu$ dependent rates:
$n + \bar e \leftrightarrow p + \bar \nu_e,\ n + \nu_e \leftrightarrow p + e, \
n  \leftrightarrow p + e + \bar \nu_e$
evolved down to the deuterium `bottle neck' temperature $T_\gamma = 0.07$ MeV obtained 
by utilizing the solution of
Eq.(\ref{1yay},\ref{2yay}) and Eq.(\ref{3yay})\footnote{
BBN implications of mirror dark matter without kinetic mixing, but with non-zero $T'/T$
arising from assumed initial conditions was studied in ref.\cite{plep}.}.
In this way we can compute the helium mass fraction for a particular value of $\epsilon$, $Y_p (\epsilon)$.
For $\epsilon = 0$, the result is the standard value, $Y_p (0) \simeq 0.24$.
We define $\delta N_{eff} [BBN]$ in the usual way:
\begin{eqnarray}
\delta N_{eff} [BBN] = {Y_p (\epsilon) - Y_p (0) \over 0.013} \ .
\end{eqnarray}
Thus $\delta N_{eff} [BBN] = 1$ corresponds to an increase of $Y_p$ of 0.013
which is equivalent to having 4 neutrinos instead of 3 contributing to the energy density.
In figure 3 we plot our results for $\delta N_{eff} [BBN]$ versus
$\epsilon$.  Comparison of figure 3 with figure 2 shows that we expect
$\delta N_{eff} [CMB] > \delta N_{eff} [BBN]$. The diminished effect for BBN is
simply because a significant part of the generation of $T'_\gamma$ and reduction of $T_\gamma/T_\nu$
occurs below the freeze out temperature of weak interactions $T_w \approx 0.8$ MeV.

\vskip 0.5cm
\centerline{\epsfig{file=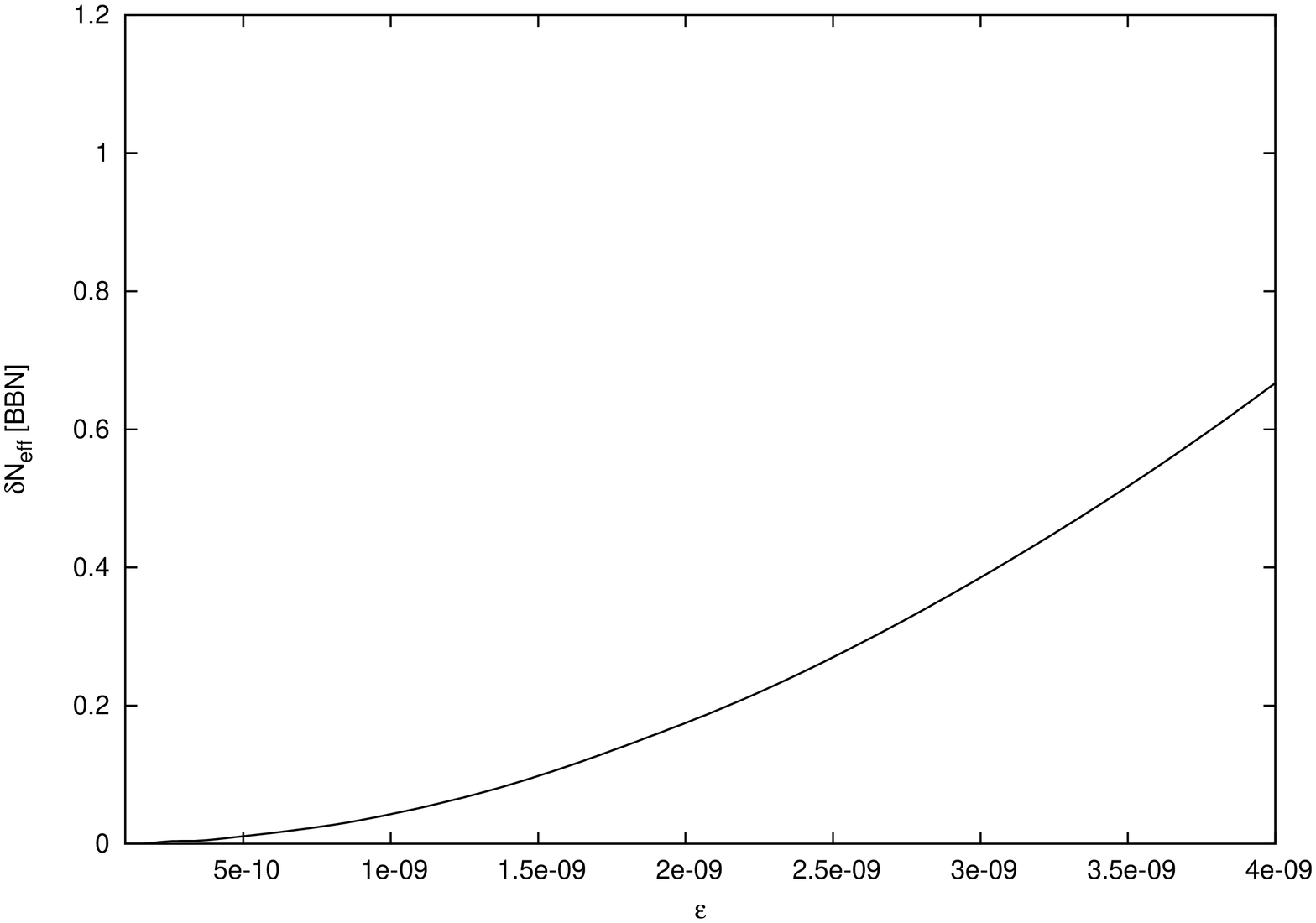,angle=270,width=13.5cm}}
\vskip 0.3cm
\noindent
{\small
Figure 3: $\delta N_{eff} [BBN]$ versus $\epsilon$.
}

\vskip 0.8cm

In conclusion, we have computed the additional radiation content of the Universe 
at the recombination epoch
due to the effects of photon-mirror photon kinetic mixing, $\delta N_{eff} [CMB]$.
We have also calculated the effects for big bang nucleosynthesis, $\delta N_{eff} [BBN]$.
There are some interesting hints that both $\delta N_{eff} [CMB]$ and $\delta N_{eff} [BBN]$ are 
non-zero and positive, which if confirmed by e.g. the anticipated high precision measurements
of the CMB from the Planck mission, will ultimately allow $\epsilon$ to be determined.

\vskip 1.5cm
\noindent
{\large Acknowledgments}

\vskip 0.2cm
\noindent
This work was supported by the Australian Research Council.

\end{document}